# Catalog of Stars with Solar-Type Activity — CSSTA[1]

Aleksey A. Shlyapnikov
aas@craocrimea.ru

## Introduction

The first lists of flare red dwarfs of UV Cet type appeared in the middle of the past century have comprised 2–3 dozen stars each. The first catalog of such stars was compiled at the Crimean Astrophysical Observatory and presented at the meeting of the International Astronomical Union in 1971. It comprised 53 objects and was placed in the VizieR database at number II/55 (Shakhovskaya, 1971).

At the end of the century, Hawley et al. (1996) carried out a spectral classification of about 2000 M dwarfs that are close to the Sun and found that 105 of them were emission M0–M3 dwarfs and 208 were M4–M8 stars with emission. Taking into account these and other results of those years in the Crimea, the GKL99 catalog was compiled (Gershberg et al., 1999) comprising 462 flare UV Cet stars and related objects in the vicinity of the Sun. In the VizieR database, the catalog is designated as J/A+AS/139/555, whereas in the search system SIMBAD it is represented as GKL99.

The GKL99 catalog was the basis for compiling a new list of stars with solar-type activity. A change of the name from "flare UV Cet stars" into "stars with solar-type activity" has marked certain progress in understanding the physics of activity. This was the GTSh10 catalog comprising 5535 objects. A detailed description of the GTSh10 catalog is given in Issue 1, V. 107 of the *Izvestiya Krymskoi Astrofizicheskoi Observatorii* (Gershberg et al., 2011).

Supplement to the monograph of R. E. Gershberg, N. I. Kleeorin, L. A. Pustilnik, and A. A. Shlyapnikov *Physics of Middle- and Low-Mass Stars with Solar-Type Activity* (M.: FIZMATLIT, 2020, 768 p., ISBN 978-5-9221-1881-1) provides a description of the second version of the catalog of stars with solar-type activity prepared at CrAO in 2019. This catalog has already included 29046 objects.

Below, we provide a description of the third version of the Catalog (CSSTA-3), present its structure and filling with data as at December 03, 2023.

## 1. Input Catalog

As in the previous version of the Catalog of Stars with Solar-Type Activity, the input list based on which the objects were selected is GAIA[2] Data Release 2 (GAIA DR2) (GAIA collaboration, 2018).

Interest to the project data in the context of compiling the current Catalog is caused by the following information contained in GAIA DR2. This is the two-color photometry of objects $G_{BP}$ (3300–6800 Å) and $G_{RP}$ (6300–10500 Å) of more than $1\times10^{9}$ objects, the detailed characteristic of light curves (∼ 400 thousand objects), effective temperature (> 160 million objects), interstellar extinction (> 87 million), color indices (> 87 million), radii and luminosities (> 76 million).

---

[1] Supplement for book: Gershberg, R.E.; Kleeorin, N.I.; Pustilnik, L.A.; Airapetian, V.S.; Shlyapnikov, A.A. "Physics of mid- and low-mass stars with solar-type activity and their impact on exoplanetary environments", Simferopol: LLC 'Forma", 2024. – 764 p., ISBN 978-5-907548-55-8

[2] GAIA – https://www.esa.int/Science_Exploration/Space_Science/Gaia

The basic criterion for including stars into the CSSTA-3 catalog was a detection of at least one of characteristic phenomena of solar activity — sporadic flares, cool spots, chromospheric emission of hydrogen and ionized calcium, X-ray and radio emission, and their location in the lower main sequence (Fig. A1).

The stars from GAIA, part of which was included into CSSTA-3, should thus meet the following criteria: $T_* < 7000°$ K, i.e., the stars should be of spectral type F5 and cooler (the left vertical border of the hatched region), $L_* < 1.1\ L_\odot$ (upper border), $L_* \geq 6.136\times10^{-6}\times T_* - 0.022$ (the equation that describes the cutting off the stars which fall into the region of white dwarfs — the left oblique border). There are 21321430 such stars in the GAIA DR2 catalog. Such a significant dataset can serve for searching for new objects with solar-type activity; however, in our case, it is used as an input catalog only and on its basis a selection of stars is implemented.

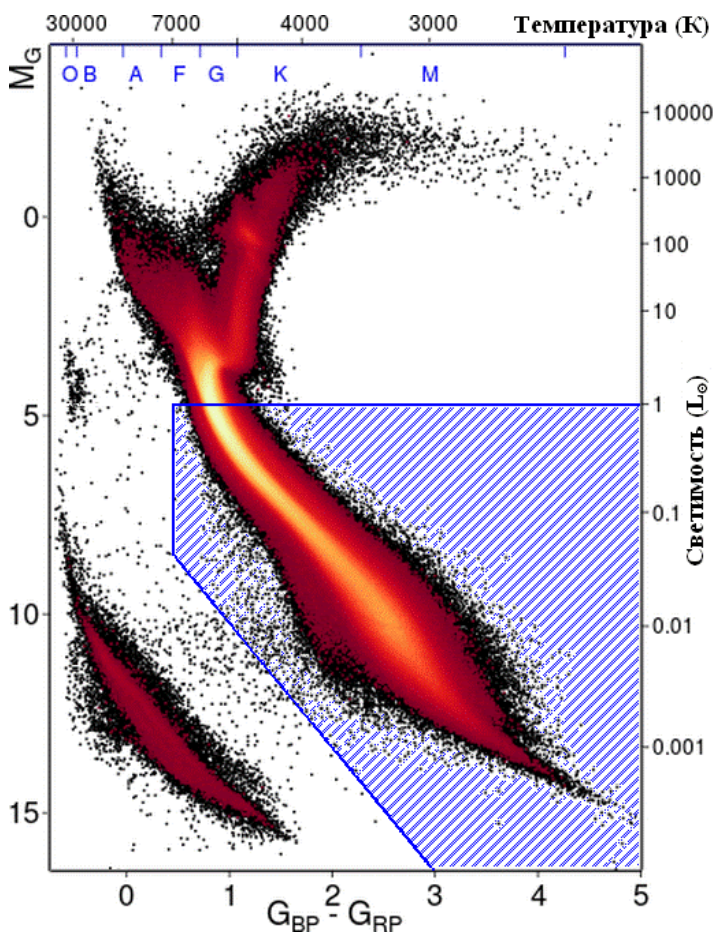

Fig. 1. A marked hatched fragment of the Hertzsprung-Russell diagram inside which the stars for CSSTA-3 were selected

With the aim of maximal using of data from the ground-based and space reviews, the input catalog was limited by $17^m$ and comprised a list of 13554156 objects.

## 2. Structure of CSSTA-3

CSSTA-3 contains 33 columns whose description is given in Table 1.

*Table 1*

| Column | Bytes | Format | Units | Label | Explanations |
|---|---|---|---|---|---|
| 1 | 1- 6 | I6 | --- | Record | Record from CSSTA-3 |
| 2 | 8- 16 | F9.5 | deg | RAJ(2000) | Right ascension (J2000) |
| 3 | 18- 26 | F9.5 | deg | DEJ(2000) | Declination (J2000) |
| 4 | 28- 64 | A37 | --- | SIM/Cat_ID | Main designation of the object according to SIMBAD or according to the source catalog |
| 5 | 66- 74 | A9 | --- | SIMBADTyp | Basic type from SIMBAD |
| 6 | 76- 79 | A4 | --- | VSXT | Variability type, as in the GCVS catalog |
| 7 | 81- 86 | F6.3 | mag | VmagSI | V magnitude from SIMBAD |
| 8 | 88- 93 | F6.3 | mag | VmagGA | V magnitude from GAIA (1) |
| 9 | 95-100 | F6.3 | mag | VmagSD | V magnitude from SDSS (2) |
| 10 | 102-107 | F6.3 | mag | maxVSX | ? Magnitude at maximum |
| 11 | 109-114 | F6.3 | mag | minVSX | ? Magnitude at minimum |
| 12 | 116-121 | F6.3 | mag | difVSX | Amplitude of variability from VSX (3) |
| 13 | 123-125 | A3 | --- | Pab | Passband from VSX |
| 14 | 127-129 | A3 | --- | Fl | Detected flares |
| 15 | 131-143 | A13 | --- | SIMBAD_Sp | Spectral type from SIMBAD |
| 16 | 145-149 | A5 | --- | GASpT | Spectral type from GAIA (4) |
| 17 | 151-153 | A3 | --- | SDS | Spectral type from SDSS |
| 18 | 155-160 | A6 | --- | W_emis | Lines emission (5) |
| 19 | 162-163 | A2 | --- | CA | Presence of the index of Chromospheric Activity (S) |
| 20 | 165-167 | A3 | --- | Spo | The presence of Spots (Spo) or their parameters (Spa) |
| 21 | 169-170 | A2 | --- | XF | X-ray emission (X) or registered flares (XF) |
| 22 | 171-172 | A2 | --- | UV | Ultraviolet radiation (6) |
| 23 | 174-175 | A2 | --- | IR | Infrared radiation |
| 24 | 177-178 | A2 | --- | RF | Radio emission (R) or registered flares (RF) |
| 25 | 181-187 | F7.2 | | Teffect | Stellar effective temperature |
| 26 | 189-192 | F4.2 | solRad | Rsol | Estimate of radius |
| 27 | 194-198 | F5.3 | solLum | Lumin | Estimate of luminosity |
| 28 | 200-211 | F16.10 | d | P_VSX | Period of a variable in days from VSX |
| 29 | 213-224 | F16.10 | d | P_KELT | Period of a variable in days from KELT |
| 30 | 226-237 | F16.10 | d | P_Kepler | Period of a variable in days from Kepler |
| 31 | 239-245 | F7.4 | y | CyclPer | Period of cyclic variability per year |
| 32 | 247-248 | A2 | --- | Pl | Presence of exoplanets (Y) and their number (Yn), or candidates (Y?) |
| 33 | 250-253 | I4 | --- | Note | Note |

Note (1): Calculated V magnitude from GAIA.
Note (2): Calculated V magnitude from SDSS.
Note (3): Difference in V magnitudes from maximum to minimum.
Note (4): Calculated spectral type from GAIA.
Note (5): Emission in the $H_\alpha$, $H_\beta$, $H_\gamma$, $H_\delta$ or CaII-K (C) lines. The + sign corresponds to the strongest line.
Note (6): Ultraviolet radiation (UV): far-UV (F), near-UV (N), far- and near-UV (FN).

## 3. Data from SIMBAD in CSSTA-3

To include the SIMBAD (Wenger et al., 2000) data into this Catalog, a series of selection criteria for objects was used according to their type, which emphasizes their physical nature. Thus, all the previous versions of the prepared at CrAO analogous catalogs included flare UV Cet stars and BY Dra variables. According to the SIMBAD data, 2259 objects were attributed to the Fl* (UV Cet) stars in 2019. In the latest version of the classification of objects in SIMBAD[1] in 2021, the stars of such a type were attributed to eruptive (!?). 1133 objects (1008 in 2019) are currently attributed to BY* (BY Dra), which are also included into the catalogs prepared earlier. Some of the indicated above stars have several determinations of variability types; among them there are T Tau, RS CVn, and other objects. All of them were deleted in the process of subsequent compilation.

After the cross identification based on coordinates in a radius of 1 arcsec of the input catalog with the SIMBAD database, 287462 objects were included into the preliminary list for further use. After excluding stars with main types that did not correspond to the required ones, 264611 objects left in the list. Taking into account that for further work with the current Catalog the single stars are preferred rather than binary or multiple stars, during the subsequent data filtration these objects were singled out into a separate list. As a result, we obtained a file containing 255191 objects.

As a final filtration of the given section of CSSTA was the singling out, into a separate list, of objects with unambiguous representation of luminosity classes, following the information from SIMBAD as compared to those determined from the GAIA data. After excluding these objects, the resultant file for the subsequent compilation into the Catalog comprises 252930 stars.

## 4. Index Catalog of Variable Stars, Types and Amplitudes

By the end of July 2021, the VSX[2] database (Watson et al., 2006) comprised information on 2114232 variable stars; among them there were 86379 BY Dra stars, 2003 UV Cet stars, and 59358 stars with detected rotational modulation.

To include into CSSTA-3, a series of selection criteria of objects was used according to their type, which emphasizes their physical nature. Thus, all the previous versions of the prepared at CrAO analogous catalogs comprised flare UV Cet stars and BY Dra variables, and all stars of these types from VSX were included into the current version of the Catalog, ruling out the objects that had several determinations of variability types and being involved into the binary or multiple systems. Particular attention was given to stars with detected rotational modulation.

After the cross identification based on coordinates in a radius of 1 arcsec of the input catalog with the VSX database, 14938 objects were involved into a preliminary list for further use. After excluding 767 stars of the main types that did not correspond to the required ones, a list prepared for CSSTA-3, with the VSX data included, comprised 252163 objects.

---

[1] SIMBAD – http://simbad.cds.unistra.fr/simbad/

[2] VSX – https://www.aavso.org/vsx/index.php

## 5. Spectral Data for CSSTA-3

At the end of March 2019, Mamajek[1] published an updated version of the database of color indices and effective temperatures for dwarf stars of the lower main sequence. The database described earlier in Pecaut et al. (2012), Pecaut and Mamajek (2013) was used during the compilation of CSSTA-3 to determine spectral types of some objects.

The database is constructed on the basis of the independent review of literature and information from different catalogs and reflects the modern system of the Morgan–Keenan (MK) spectral classification. It represents the spectral types of dwarfs from O3V to Y2V by the effective temperature, bolometric luminosity (normalized to the solar one), bolometric stellar magnitude, bolometric correction in the V band, absolute stellar magnitude in the V band, and corresponding color indices.

Color indices are based on the photometry in the Johnson *U*, *B*, *V* and Cousins $R_C$, $I_C$ systems, the Tycho catalogs of *Bt*, *Vt*, and *G*, *Bp*, *Rp* of GAIA DR2, data on *i*, *z*, *y* from Sloan (SDSS), *J*, *H*, *K* from 2MASS, and *W1*, *W2*, *W3*, *W4* from WISE projects.

Checking consistency of spectral types from the literature sources was implemented based on the optical observations of stars of spectral types from F3 to M4 carried out with the 1.5-meter telescope SMARTS mounted on the Sierra del Toro in Chile.

When refining and determining the spectral types of objects from CSSTA-3, the General Catalog of Stellar Spectral Classification (Skiff, 2014) was used containing 888588 bibliographic references to the spectral classification of 660614 stars. The number of stars later than F5V is 45547.

The previous version of the Catalog comprised a sample of information on the spectroscopy of M dwarfs from the SDSS Data Release 7 (West et al., 2011). To obtain information, the 2.5-meter wide-angle telescope with a multi-aperture spectrograph was used belonging to several universities of the USA, which is located at the Apache Point Observatory (New Mexico, USA).

In 2011, on the basis of data of the 7th implementation of the SDSS project, the catalog of spectral observations of 70841 M dwarfs was published. It involved five-color photometry in the SDSS system, determinations of spectral types of stars, data on the $H_\alpha$, $H_\beta$, $H_\gamma$, $H_\delta$, and CII-K lines and other information. After their detailed analysis, these data were partially integrated into CSSTA-3.

After the cross identification based on coordinates in a radius of 1 arcsec of the input catalog with the SDSS database, 63783 objects were involved into a preliminary list for further use. After excluding 721 objects of main types that did not correspond to the required ones and 600 binary objects from the SDSS database, 62462 objects were included into a list prepared for the current Catalog.

## 6. Observational Data from the Kepler Orbital Station

Kepler[2] is a space observatory of NASA launched in 2009 with the aim of searching for planets that orbit stars. Including the K2 project as a result of the mission, 3032 candidates for exoplanets were recorded; among them 3253 had been confirmed by mid-November 2022. Data of the Kepler project were of interest for the current version of the Catalog in two directions. First, the search for flare activity and, second, the presence of exoplanets in dwarf stars of the lower main sequence.

---

[1] Mamajek, 2019 – http://www.pas.rochester.edu/~emamajek/EEM_dwarf_UBVIJHK_colors_Teff.txt
[2] Kepler – https://www.nasa.gov/mission_pages/kepler/main/index.html

A homogeneous search for stellar flares was carried out using all the available light curves obtained from Kepler. Within the first observing set of Kepler in 2011, 23000 K and M dwarfs were investigated. Thus, 373 flaring stars were identified; some of them exhibited several events throughout the whole observational period (Walkowicz et al., 2011). In 2012, Maehara et al. (2012) reported on the observation of 8300 stars from Kepler throughout 120 days and on the detection of 148 G stars for which flares with energies of $10^{29}$ to $10^{32}$ erg were recorded on time intervals of hours. A 2016 publication describes 851168 random outbursts detected from Kepler observations of 4041 stars (Davenport, 2016). However, a revision of Davenport's data was reported in 2019, which indicates that his list includes various pulsating stars, rapidly rotating stars, and stars with transit events that can cause either flares or rapidly changing brightness, leading to false positive observational processing signals (Yang and Liu, 2019). Comparison of Davenport's catalogs with those of Yang and Liu showed that only 396 stars from the first list are contained in the second one.

In 2017, Van Doorsselaere with colleagues published information on the detection of 16850 flares on 6662 stars observed by the Kepler observatory (Van Doorsselaere et al., 2017). A cross identification of the list with the data of Yang and Liu has already ensured the coincidence of 2223 objects. The rest objects, in the opinion of the Catalog's authors, being compared to the data of Van Doorsselaere et al., were burdened with the same shortcomings as Davenport's catalog.

## 7. Observational Data from the TESS Orbital Station

The orbital observatory for performing surveys of the sky with the aim of searching for exoplanets by the transit method (Transiting Exoplanet Survey Satellite — TESS[1]) analyses the brightness of bright and nearby stars (Ricker et al., 2014).

A number of papers are devoted to the analysis of observations from TESS, in particular, to the search for flare activity of red dwarfs. Let us consider a procedure for complementing CSSTA-3 with the data of these studies on the example of two publications "Stellar flares from the first TESS data release: exploring a new sample of M dwarfs" (Günther et al., 2020) and "A catalog of cool dwarf targets for the Transiting Exoplanet Survey Satellite" (Muirhead et al., 2018).

The first publication represents observations of 24809 stars with a 2-min exposure during the first two months of the TESS mission. The authors identified 1228 flare stars, 673 of them were M dwarfs. The final list involved 764 stars that exhibit flare activity; 632 of them were M dwarfs.

The results of these studies are placed in the VizieR database and represented by two tables: the catalog of individual flares found from the observational data of TESS in the first and second sectors (8695 flares) and the catalog of flare stars found by authors (1228 objects).

Among objects of the second table, 1131 are designated as dwarfs, 39 — as giants, and 58 objects have an undetermined luminosity class. To include into CSSTA-3, a cross identification of objects from the second table with the current version of the Catalog was carried out. CSSTA was supplemented with information on flares on 60 stars (Gorbachev, 2023).

---

[1] TESS – https://www.nasa.gov/tess-transiting-exoplanet-survey-satellite

## 8. The KELT Project and Stellar Rotation Periods for CSSTA-3

A very small "thousand-degree" telescope[1] consists of two instruments whose optics includes 80 mm objectives Mamiya 645 (f/1.9 with an effective aperture of 42 mm). The field of view of telescopes on the celestial sphere covers an area of 26°×26°. Instruments are located at the Winer Observatory in Sonoita, Arizona (USA) and at the South African Astronomical Observatory near Sutherland (Oelkers et al., 2018).

For almost 10 years of observations within the project, there was recorded information on 4000000 sources in the range of magnitudes $7^m < V < 13^m$ located in more than 70% of the sky. The basic scientific objective of the study is a detection of transit phenomena during the passage of planets with large radii on the background of bright host stars. In 2018, there was published a catalog of observations carried out within the KELT project containing information on 52741 objects which exhibit significant brightness variations of high amplitude probably caused by variability, including flaring one, as well as data for 62229 objects identified with probable stellar rotation periods. In CSSTA-3, the information from the KELT project is used to involve flare stars and data on rotation periods.

In the VizieR database, the results of observations within the KELT project are represented by three tables: 1 — contains astrometric information, information on stellar magnitudes and brightness variations for variables from KELT; 2 — upper limits of variability for non-varying sources in the TESS Input Catalog (3873790 objects), and 3 — information on the periodic brightness variations for the dwarf candidates from the TESS Input Catalog (TIC). To include into the current version of the Catalog, a cross identification of objects from the third table of the KELT catalog with the Input Catalog was performed.

The cross identification resulted in a preliminary list of 7022 stars from the KELT catalog to include into CSSTA-3. The list format corresponds to the structure of this Catalog according to the sections involved into the compilation.

The inclusion of information from the KELT database into CSSTA-3 made it possible to perform a comparative analysis of stellar rotation periods determined from the VSX and KELT data. It should be noted a significant number of periods from the KELT data that are close to one day, which is obviously a result of their automatic search and is not true. The periods presented in VSX were determined in the process of observational data reduction, which makes them more reliable. The problem of "similar periods" for a significant number of objects is well illustrated as the horizontal lines overlapping the whole range of effective temperatures, radii, and luminosities. The probability that tens and hundreds of stars have similar rotational periods seems to be very doubtful.

## 9. Supplementation of CSSTA-3 with the GTSh10 Data

After creating GTSh10[2], there appeared the data which allowed us to revise information contained in this catalog. The Catalog of Stars with Solar-Type Activity designated as GTSh10 was prepared in 2010. The data from publications of the preceding 10–15 years were used for its compilation. The Catalog mainly included dwarf stars with different manifestations of solar-type activity. It was composed of objects with dark spots, hydrogen and calcium chromospheric emission, fast flares in different wavelength ranges, and with radio and X-ray emission of stellar coronae. The compiled list comprises 5535 objects.

---

[1] KELT — https://keltsurvey.org/telescopes
[2] GTSh10 — http://www.crao.ru/~aas/CATALOGUEs/G+2010/eCat/G+2010.html

To include into CSSTA-3, a cross identification of the objects from GTSh10 with the compiled catalog was performed. As a result of the cross identification with a radius of 1 arcsec, 4733 objects from GTSh10 were found among the objects of the CSSTA-3 Input Catalog, whereas 802 objects were absent.

Since the cross identification was performed based on the coordinates, whereas most objects from GTSh10 have significant proper motions, then it is possible that for 802 stars the coordinates were specified with insufficient accuracy.

For 169 objects of the cross identification, according to the GAIA data, there were no values of luminosity and stellar radius, whereas for 171 — no values of temperature. Two objects have a temperature of more than 8000 K. 1444 stars have radii and luminosities in solar units exceeding those specified when compiling the Input Catalog. Taking into account that objects were included into GTSh10 based on the results of the analysis of publications with their detailed description, after a closer examination, all the stars that correspond to the specified criteria were involved into CSSTA-3.

## 10. X-Ray Sources for CSSTA-3

This section presents the identification of objects from CSSTA-3 in the X-ray wavelength range including that for objects of the optical range in the hard emission regions in proximity to red dwarf stars.

Various catalogs and databases were used to include information on the presence of X-ray emission in stars into the Catalog. In the previous version of CSSTA-3, 2909 objects with registered X-ray fluxes were presented. Independent processing of the "first light" image of the eROSITA telescope (Predehl et al., 2021) aboard the SRG observatory (Sunyaev et al., 2021) made it possible to identify 2485 X-ray objects.

Taking into account that the pixel size in the working image is ~ 15", the search for candidates for identifying isolated X-ray sources with optical objects was carried out in the area less than 40" to avoid edge processing effects. After the cross-identification of identified X-ray sources with the GAIA DR2 catalog, a list of matches amounted to 2868 objects. It is obvious that when the size of the search for GAIA DR2 objects in areas with a radius of < 20", there are non-single coincidences. In this case, up to 6 GAIA DR2 objects fell into the X-ray emission region. In order to obtain a one-to-one correspondence between X-ray sources and red dwarfs, regions with non-single coincidences were excluded from further consideration. After their exclusion, 899 objects remained in the list, 67 of which meet the criteria for temperature, radius, and luminosity for including into the Catalog (Shlyapnikov, 2021a).

The study of the equatorial region of the sky (eFEDS — eROSITA Final Equatorial Depth Survey) became the longest observation in the period of testing the capabilities of the eROSITA telescope instruments. In total, about 100 hours were spent. The eFEDS field has an area of approximately 140 deg$^2$ and consists of four separate rectangular scan areas of 35 deg$^2$. The site was chosen due to the presence of a significant number of multiwave observations of this region of the sky.

Based on the results of the eFEDS deep survey, two catalogs of X-ray sources were published: unambiguously detected sources in the 0.2–2.3 keV range and hard (2.3–5 keV) sources observed in the multiband mode (Brunner H. et al., 2022 ). To identify objects from CSSTA, 27910 X-ray sources with a high degree of detection probability were selected, which make up the main catalog of eFEDS.

Taking into account that the average radius of errors in determining the eFEDS coordinates is 4.8 arcsec, the search for identifying candidates was carried out in a threefold radius. As a result, among the objects of a deep survey of the equatorial region of the sky, the stars that fall into 110 fields limited by the radius of errors in determining X-ray coordinates were identified. Twelve stars were previously classified as X-ray sources, which confirms the correctness of independent identification. Galaxies were found in the identification regions of two stars, one of which is a known X-ray object. Several identities contain closely spaced objects. All areas were analyzed visually, and the presence of X-ray radiation was indicated in the Catalog for objects that met the selection criteria (Shlyapnikov, 2021b).

The previous version of CSSTA contained data on the registered X-ray emission from 2909 objects. To supplement the Catalog with information on stellar X-ray emission, the MORX compilation catalog (Flesch, 2023) was used. It included data related to the observation of objects in the X-ray range by the XMM-Newton[1], ROSAT[2], Chandra[3], and Swift[4] observatories.

In total, the catalog contains 3115575 optical objects. Each object has coordinates for epoch 2000, optical and radio/X-ray identifiers, photometry in the red and blue regions of the spectrum, and values of the calculated match probabilities.

The description of the MORX catalog contains information about the classification of objects made by the author. In total, 18 types of sources were classified. Among them there are extragalactic (actual galaxies, galaxies with active nuclei, objects of the BL Lac type, quasars and others) and galactic (star formation regions, cataclysmic variables, white dwarfs), including objects of an unknown type but with a predetermined redshift according to the SDSS data.

For analysis and subsequent inclusion into CSSTA, out of 3115575 MORX objects, 1357332 objects were selected that have classification as stars, X-ray or radio sources, as well as sources of unknown type. For the purpose of independent identification in the optical range of the spectrum, as well as extracting information about the proper motions of objects (to exclude possible extragalactic sources), the cross-identification of selected objects with stars from the GAIA DR2 catalog was performed.

After cross-identification with a radius of 5 arcsec (based on the fact that the PSF of the image of a point optical object should be at least 3 arcsec at a level of the half-width of approximation, and the search is performed in a region with a triple radius), out of 1357332 MORX sources in the GAIA DR2 catalog, 525337 objects were detected.

Considering that a significant number of the GAIA DR2 objects have the same coordinates but are not close pairs, they were excluded from further consideration. Note that most of these objects lack information about the effective temperature, radius, and luminosity. Objects that did not meet the selection criteria were also removed. As a result, the number of remaining objects amounted to 73031.

For identification by visual control, a special interactive interface was developed that allowed opening the area of the object under study in a new browser window with mapping information from the SIMBAD and NED[5] databases. The need for visual control was due to the presence of a large number of galaxies near some stars, one of which could be a source of X-rays.

Note that most of the objects under consideration are poorly studied, and there is no information about them in SIMBAD.

---

[1] XMM-Newton – https://www.cosmos.esa.int/web/xmm-newton
[2] ROSAT – https://www.mpe.mpg.de/ROSAT
[3] Chandra – https://chandra.harvard.edu
[4] Swift – https://www.nasa.gov/mission_pages/swift/main
[5] NED – https://ned.ipac.caltech.edu/

After monitoring all 73031 objects, only those objects that uniquely correspond to stars were added to CSSTA. At the beginning of November 2023, the cross-identification of CSSTA and selection from MORX gave a match for 4448 objects. Among them, there are 1371 stars with X-ray emission and 3077 X-ray sources identified with stars.

## 11. Ultraviolet Sources for CSSTA-3

The basic information on ultraviolet radiation for objects from the current Catalog was acquired through the cross identification with the observational data of the GALEX (Galaxy Evolution Explorer)[1] observatory. The NASA mission was started on April 28, 2003 and operated until June 28, 2013. The observatory was equipped with the 50-cm Ritchey-Chrétien Telescope (f/6.0) with a field of view of 1.2° operating in two ultraviolet ranges, far (FUV, $\lambda_{eff} \sim 1528$ Å) and near ultraviolet (NUV, $\lambda_{eff} \sim 2310$ Å). The GALEX database contains FUV and NUV images, about 500 million measurements, and more than 100000 low-resolution ultraviolet spectra.

For the cross identification of CSSTA-3 with the GALEX data, an updated version of the catalog of ultraviolet sources was chosen (Bianchi et al., 2017). The catalog includes all observations from the survey with the largest coverage of the sky regions recorded with both FUV and NUV detectors. This is more than 28700 fields and, in total, 57000 observations. The catalog is composed of two parts that contain 82992086 and 69772677 sources with the typical penetrating value in FUV = $19^m.9$ and NUV = $20^m.8$. The second part of the catalog is a more limited version that uses only a central part of each observed field of 1°.

As a result of the cross identification with a radius of 1 arcsec, 105542 stars were detected among the GALEX sources; the information on them was included into CSSTA-3.

## 12. Radio Sources for CSSTA-3

To supplement CSSTA-3 with information about the radio emission of stars, the MORX catalog described in Section 10 was used. Earlier, CSSTA indicated the registration of radio emission from 95 stars. The MORX catalog contains information obtained in the radio band in the NVSS[2], FIRST[3], SUMSS[4], and other projects. Cross-identification based on the coordinates of MORX objects and stars from CSSTA yielded a match for 361 objects.

## 13. Other Additions to CSSTA-3

### 13.1. Flares on Red Dwarf Stars

The detailed analysis of red dwarf stars with manifestations of flare activity included into CSSTA-3 is described in Sections 6 and 7. Since information on the studies was published before 2018, then by the moment of preparing the current version several new papers had been published and their data were added into CSSTA-3.

---

[1] GALEX – http://www.galex.caltech.edu/
[2] NVSS – https://www.cv.nrao.edu/nvss/
[3] FIRST – http://sundog.stsci.edu/
[4] SUMSS – http://www.astrop.physics.usyd.edu.au/sumss/

## 13.2. Cycles of Lower Main-Sequence Stars

### 13.2.1. Long Cycles

Monitoring of chromospheric activity yields valuable information on the stellar magnetic activity and its dependence on the fundamental parameters, such as effective temperature and rotation. Boro Saikia et al. (2018) represent a catalog of chromospheric activity of 4454 cool stars based on the combination of archival HARPS spectra and a number of other surveys including the Ca II H&K data from the Mount Wilson Observatory project. On the basis of this work, the current Catalog was supplemented with the mean S indices of chromospheric activity for 2559 objects with an identification radius of 1 arcsec and 53 data on the detected rotation periods and cycles.

### 13.2.2. Short Cycles

Along with studying long activity cycles over the past decades, owing to such projects as Kepler and TESS, there appeared a possibility of measuring cyclic variations of the stellar light curve amplitude and rotation period on timescales of a few years. Thus, using the Kepler data for four years (Reinhold et al., 2017), 23601 stars were studied. The periodic amplitude in the 0.5–6 year interval with a rotation period of 1–40 days was detected for 3203 stars.

The performed cross identification in a radius of 1 arcsec with the CSSTA-3 data made it possible to detect 874 objects involved into the new Catalog.

## 13.3. Exoplanets

Information about stars with confirmed or suspected exoplanets was added to the catalog. Analysis of the presence of X-ray and radio emission, if any, should contribute to understanding the possibility of the existence of "life" in the habitable zone. Flare activity also imposes certain restrictions on the development of life. Changes in brightness associated with the orbital periods of exoplanets around stars are superimposed on the overall light curve of a star. They must be taken into account when analyzing rotation and the period of possible cyclic activity.

The inclusion of information on exoplanets into the catalog was preceded by their studying at CrAO, which has been conducted since 2000. A detailed description of the studies is presented on the website[1].

By mid-November 2023, according to the NASA Exoplanet Archive[2], it had been confirmed the existence of 5550 planets discovered by 11 methods and 410 found by the TESS observatory. The discovery of 6977 possibly existing exoplanets according to TESS observations needs to be confirmed. Among the detection methods, the leader is the observation of the passage (transit) of the planet against the background of a star — 4176 discoveries. 1070 planets were discovered by changing the radial velocities of lines in the spectra of stars. 204 planets were detected by using the microlensing method. From 1 to 69 planets were found by using other methods.

---

[1] CrAVO exoplanet – https://sites.google.com/view/cravo-exop-psa
[2] NASA exoplanet archive – https://exoplanetarchive.ipac.caltech.edu

After cross-identification by coordinates between the Catalog of Confirmed Planets (CCP) and the Data Base of Candidates (DBC), it was found that 945 stars from the CCP and 771 from the DBC fall into CSSTA.

An analysis of the distribution of stars from the CCP by magnitude showed that the maximum distribution falls on the range from $12^m$ to $13^m$. This is due to the use of short-focus lenses for panoramic surveys of the sky in order to search for transits of exoplanets, as well as long-focus instruments when searching for changes in radial velocities. In the first case, restrictions are imposed on the penetrating power of astrographs, and, in the second case, on threshold restrictions on the signal-to-noise ratio in spectroscopy.

The maximum in the distribution of spectral types of stars falls on K1. Note that this sample of objects identified by CSSTA does not contradict the spectral classification of all stars with exoplanets described in Gorbachev et al. (2019), where the first maximum of the distribution is also in the region of spectral types K.

When analyzing the distributions of the periods of variability of the considered stars with exoplanets according to the VSX data and the orbital periods of planets around stars, it was recorded that the maxima in both distributions fall on 3 days. At the same time, it should be noted that 25 objects have the type of variability BY Dra according to the SIMBAD database, and 467 — RotV*, i.e., in the first and second cases, their variability is caused by the rotational modulation of the spotted surface rather than by the presence of exoplanets. Considering that the brightness variation amplitude for these stars lies in the range from $0^m.001$ to $0^m.1$, it is possible that most of the objects included in the Catalog are incorrectly classified according to the type of variability.

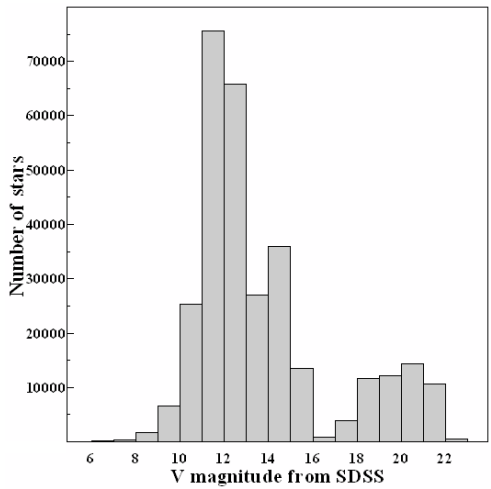

Fig. 2. Distribution of the number of stars from CSSTA-3 by the stellar magnitude V

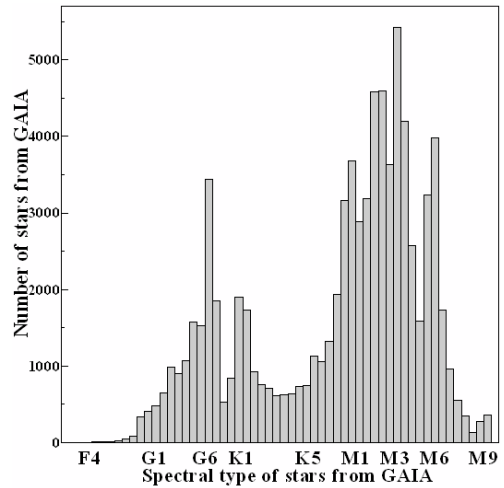

Fig. 3. Distribution of the number of stars from CSSTA-3 by spectral types

## 14. Version of CSSTA-3 as at 03.12.2023

The total number of stars in CSSTA-3 as at December 03, 2023 is 314618.

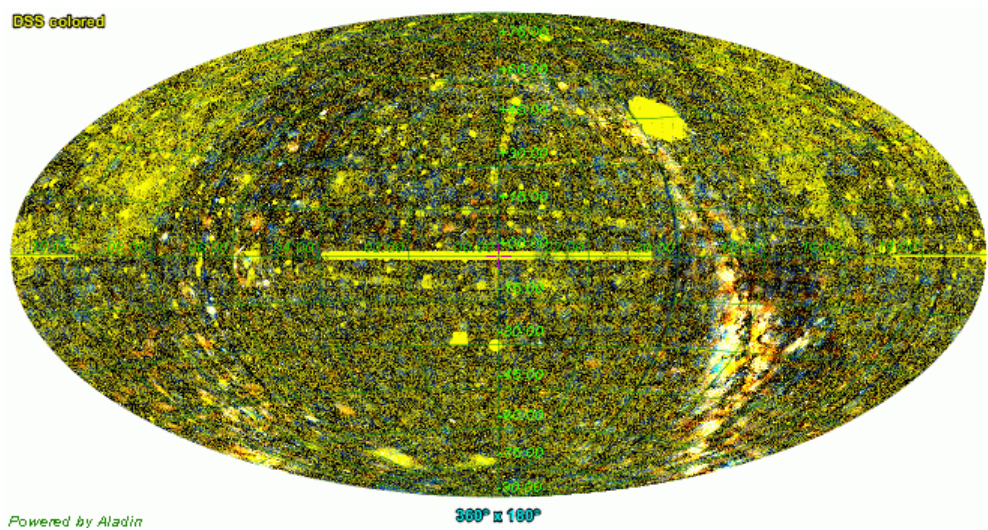

Fig. 4. Illustration of the distribution of 314618 objects from CSSTA-3 in the sky

## 15. Conclusions

After compiling all the described above sources, excluding coincidences and a number of other operations, the current CSSTA-3 catalog of the lower main-sequence stars with solar-type activity was obtained. It comprises 314618 objects, and the database that is realized on its basis is a developing project that contains hyperlinks to the original photometric and spectral observations. The Catalog is available on the website of the Crimean Astrophysical Observatory at http://craocrimea.ru/~aas/CATALOGUEs/S-2019/eCat/S-2019.html. More information about CSSTA-3 can be seen in ADS/NASA or VizieR.

To provide a wider access to the Catalog and the database of stars with solar-type activity, a copy of the information from the CrAO server is available via the Google resource at https://sites.google.com/view/csast.